\documentclass[twocolumn,showpacs,preprintnumbers,amsmath,amssymb]{revtex4}
\usepackage{graphicx}% Include figure files
\usepackage{dcolumn}% Align table columns on decimal point
\usepackage{bm}% bold math

\begin{document}

\preprint{APS/123-QED}

\title{Photon number states generated from a continuous-wave light source}

\author{Anne E. B. Nielsen and Klaus M\o lmer}
\affiliation{Lundbeck Foundation Theoretical Center for Quantum
System Research, Department of Physics and Astronomy, University of
Aarhus, DK-8000 \AA rhus C, Denmark}

\date{\today}

\begin{abstract}
Conditional preparation of photon number states from a
continuous-wave nondegenerate optical parametric oscillator is
investigated. We derive the phase space Wigner function for the
output state conditioned on photo detection events that are not
necessarily simultaneous, and we maximize its overlap with the
desired photon number state by choosing the optimal temporal output
state mode function. We present a detailed numerical analysis for
the case of two-photon state generation from a parametric oscillator
driven with an arbitrary intensity below threshold, and in the low
intensity limit, we present a formalism that yields the optimal
output state mode function and fidelity for higher photon number
states.
\end{abstract}


\maketitle

\section{Introduction}

The study of nonclassical states of light has provided a deeper
understanding of quantum fluctuations and the role of measurements
in quantum theory and it has led to applications in precision
metrology and quantum communication. The photon number states, or
Fock states, play a special role, because they have vanishing
intensity fluctuations, and their interaction, e.g., with a single
two-level atom in an optical cavity, is particularly regular. By
injection of atoms with properly selected excitation and passage
times through a micro maser, it is possible to build up a wide range
of states, including number states of the cavity field
\cite{walther,haroche}, and recent progress in extending the photon
lifetimes in microwave cavities \cite{harochenature} bring promise
for further experimental progress in this direction.

By control of single photo emitters such as a single molecule
\cite{molecule}, a color center \cite{colorcenter}, or a quantum dot
\cite{dot}, it is also possible to generate traveling light pulses
in the optical frequency range that contain only a single photon.
For a review on single photon emitters see \cite{singlereview}. In
these schemes, however, the production of higher number states is
not straightforward, and in the present paper we shall address an
alternative conditional approach where the signal beam from a
nondegenerate optical parametric oscillator (OPO) is projected into
the desired state by a quantum measurement performed on the idler
beam. This method has recently been used to generate single- and
two-photon states from an OPO driven by a pulsed pump field
\cite{grangier2}. Conditional generation of nonclassical states was
proposed by Dakna et al \cite{dakna}, see also \cite{kim,fiurasek},
and generation of single-photon and Schr\"odinger cat states has
been demonstrated in experiments where measurements performed on a
small fraction of the light beam from a degenerate OPO caused the
projection of the remaining beam \cite{grangiercat,jonas,wakui}.
Single-photon and cat state production from continuous-wave OPOs has
been studied theoretically in \cite{sasaki,klaus,nm}, and in the
present paper we generalize the analysis to continuous-wave
generation of higher photon number states. In contrast to single
photon and Schr\"odinger cat state production, generation of states
with two or more photons involves multiple conditioning photo
detection events, and in the continuous-wave case, we are
particularly interested in determining how the generated state is
affected by the temporal separation of the conditioning detections.
We analyze this feature generally for two-photon states and we
present an analytical treatment, restricted to the low intensity
limit, for $n$-photon state generation.

\begin{figure}
\begin{center}
\includegraphics*[viewport=5 5 155 97,width=0.85\columnwidth]{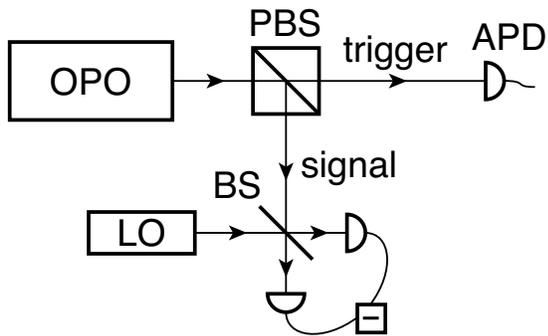}
\caption{Experimental setup for conditional preparation of Fock
states from a type II (i.e.\ polarization) nondegenerate OPO. PBS:
polarizing beam splitter, APD: avalanche photodiode, BS: beam
splitter, and LO: local oscillator.} \label{setup}
\end{center}
\end{figure}

Figure \ref{setup} exemplifies the experimental setup used to
generate Fock states. A nondegenerate OPO produces pairs of
distinguishable photons. The two kinds of photons are separated to
produce two correlated twin beams. One beam (denoted the trigger) is
observed with an avalanche photodiode (APD) detector. Detection of
$n$ close detector clicks in the trigger arm projects the state in
the signal beam into an $n$-photon state, and the generation is
verified by homodyne detection. Since the field is a continuous-wave
field the temporal mode occupied by the produced state needs to be
specified, and the largest $n$-photon fidelities are obtained by
optimizing the choice of signal state mode function. We shall
investigate to which extent the $n$ photons in the signal beam may
occupy a single mode despite the click events happening not exactly
simultaneously.

In order not to miss close clicks due to the finite dead time of
real detectors it might be advantageous to split the trigger beam
and send it onto more than one APD detector, as was done in the
two-photon experiment with pulsed fields \cite{grangier2}. However,
we show that our theoretical expression for the conditional output
state, and thus also the fidelity and the optimal signal state mode
function, is independent of the number of (ideal) APD detectors
used.

In Sec.\ II we start out with a two-mode treatment of the two-photon
state generation process. In Sec.\ III we generalize to the
multi-mode case valid for continuous-wave fields. We determine the
Wigner function for the output state conditioned on two trigger
detector click events, and we calculate the two-photon state
fidelity as a function of the signal state mode function. In Sec.\
IV we optimize the signal state mode function over all real
functions to obtain maximal two-photon state fidelity. Finally, in
Sec.\ V, we consider $n$-photon state generation in the low
intensity limit. We describe the state produced in the signal beam
in terms of photons occupying specific temporal modes, and we
determine the optimal output state mode functions. Sec.\ VI
concludes the paper.

\section{Output state conditioned on two trigger detector click
events -- two-mode treatment}

In this section we describe the two-photon state generation in the
context of a simple two-mode theory to introduce some of the basic
ideas. This treatment is approximately valid when a pulsed pump
field is used. The initial state generated by the nondegenerate OPO
is a two-mode squeezed vacuum state \cite{ban}
\begin{equation}\label{psii}
|\psi_i\rangle=\frac{1}{\cosh(r)}\sum_{n=0}^\infty
\tanh^n(r)|n,n\rangle,
\end{equation}
where $r$ is the squeezing parameter and the first (second) quantum
number inside the ket on the right hand side is the number of
photons in the trigger (signal) mode.

We assume that a trigger detector click results in the
transformation $\rho\rightarrow\hat{a}_t\rho\hat{a}^\dag_t
/\mathrm{Tr}(\hat{a}_t\rho\hat{a}^\dag_t)$ of the density operator
$\rho$, where $\hat{a}_t$ is the trigger mode annihilation operator.
We apply the click transformation twice to the state \eqref{psii},
and since we do not subject the trigger mode to further
measurements, we trace over the trigger mode afterward and
renormalize to obtain the conditional single-mode output state
\begin{multline}\label{psic}
\rho_{click}=\\
\frac{1}{2\cosh^2(r)\sinh^4(r)}
\sum_{n=0}^\infty\tanh^{2n}(r)n(n-1)|n\rangle\langle n|.
\end{multline}
The vacuum and the single-photon state contributions are eliminated
by the conditioning procedure, and the generated state is a
superposition of a two-photon state and higher photon number states.
The two-photon state fidelity is easily obtained from \eqref{psic}
as
\begin{equation}\label{fid}
F_2=\langle2|\rho_{click}|2\rangle=\frac{1}{\cosh^6(r)}.
\end{equation}
The fidelity approaches unity in the limit where the squeezing
parameter is small, because a small $r$ corresponds to a weak pump
field, and hence the probability to produce more than two photon
pairs within a single pulse is small.

For the multi-mode case it turns out to be convenient to describe
the initial unconditional state and the conditional state in terms
of Wigner functions, and we hence introduce this alternative
approach now. Since the OPO is a Gaussian light source, the two-mode
Wigner function for the initial unconditional state is a Gaussian
\begin{equation}\label{wvy}
W_V(y)=\frac{1}{\pi^2\sqrt{\det(V)}}\exp\left(-y^TV^{-1}y\right).
\end{equation}
$y=(x_t,p_t,x_s,p_s)^T$ is a column vector of quadrature variables
for the trigger and the signal mode, and $V$ is the covariance
matrix. In terms of the operators
$\hat{y}=(\hat{x}_t,\hat{p}_t,\hat{x}_s,\hat{p}_s)^T$, defined as
$\hat{x}_{t/s}=(\hat{a}_{t/s}+\hat{a}^\dag_{t/s})/\sqrt{2}$, and
$\hat{p}_{t/s}=-i(\hat{a}_{t/s}-\hat{a}^\dag_{t/s})/\sqrt{2}$, where
$\hat{a}_s$ is the annihilation operator for the signal mode, the
elements of the covariance matrix are
$V_{ij}=\langle\hat{y}_i\hat{y}_j\rangle+\langle\hat{y}_j\hat{y}_i\rangle$,
and from \eqref{psii} we obtain
\begin{eqnarray}
V_{11}=V_{22}&=&V_{33}=V_{44}=\cosh(2r),\\
V_{13}=V_{31}&=&-V_{24}=-V_{42}=\sinh(2r),
\end{eqnarray}
while the other matrix elements are zero.

The transformation from \eqref{psii} to \eqref{psic} is translated
into (see \cite{klaus,nm})
\begin{multline}\label{Wclicks}
W_{click}(x_s,p_s)=N_{click}\iint\frac{1}{4}
\bigg(1+x_t^2+p_t^2+x_t\frac{\partial}{\partial x_t}\\
+p_t\frac{\partial}{\partial p_t}
+\frac{1}{4}\frac{\partial^2}{\partial x_t^2}
+\frac{1}{4}\frac{\partial^2}{\partial
p_t^2}\bigg)^2W_V(y)dx_tdp_t\\
=\frac{1}{\pi\cosh^3(2r)}\bigg(1-2\frac{1+\cosh(2r)}{\cosh(2r)}
\left(x_s^2+p_s^2\right)\\
+\frac{[1+\cosh(2r)]^2}{2\cosh^2(2r)}\left(x_s^2+p_s^2\right)^2
\bigg)\exp\left(-\frac{x_s^2+p_s^2}{\cosh(2r)}\right),
\end{multline}
where $N_{click}$ is a normalization constant. The two-photon state
fidelity of the generated state is given by
\begin{equation}\label{F2}
F_2\equiv2\pi\iint W_{click}(x_s,p_s) W_{n=2}(x_s,p_s)dx_sdp_s,
\end{equation}
where $W_{n=2}$ is the Wigner function for a two-photon state.
Equation \eqref{F2} once again leads to the result \eqref{fid}.

\section{Output state conditioned on two trigger detector click
events -- multi-mode treatment}

In the continuous-wave case the field annihilation operators are
time-dependent and satisfy the commutator relation
$[\hat{a}(t),\hat{a}^\dag(t')]=\delta(t-t')$. In the following we
denote the trigger beam annihilation operator $\hat{a}_+(t)$ and the
signal beam annihilation operator $\hat{a}_-(t)$ to distinguish them
from the single mode operators in the last section. In principle,
there are now infinitely many modes, but since we can trace out all
unobserved modes, we only need to consider the two trigger modes, in
which the conditioning trigger detector clicks occur, and the signal
mode occupied by the generated state, which is a great
simplification. It is necessary to include two trigger modes since
the conditioning clicks may happen at different times. Initially we
assume that the trigger modes are distinct.

The temporal shapes of the relevant modes are given by the mode
functions $f_i(t)$, $i=1,2,3$, where 1 and 2 are trigger modes while
3 is the signal mode. The single-mode operators (corresponding to
$\hat{a}_t$ and $\hat{a}_s$) are then given by
\begin{eqnarray}\label{tmodeope}
\hat{a}_i&=&\int f_i(t')\hat{a}_+(t')dt', \quad i=1,2,\\
\label{smodeope}\hat{a}_3&=&\int f_3(t')\hat{a}_-(t')dt'.
\end{eqnarray}
$[\hat{a}_i,\hat{a}^\dag_i]=1$ implies that $\int |f_i(t')|^2dt'=1$.
We assume that the trigger modes are top hat functions of
infinitesimal width $\Delta t_c$ and height $1/\sqrt{\Delta t_c}$
centered at the $i$th detection time $t_{ci}$. This is valid if the
duration of a detection is much smaller than the inverse of the
leakage rate $\gamma$ of the OPO output mirror. The signal mode
function is used to specify the output state, and hence it may be
chosen arbitrarily. In Sec.\ IV we use this freedom to maximize the
two-photon state fidelity. We assume throughout the two-photon state
analysis that the signal mode function is real. Imperfect detection
may be taken into account by replacing $\hat{a}_\pm(t)$ with
$\sqrt{\eta_{t/s}}\hat{a}_\pm(t')
+\sqrt{1-\eta_{t/s}}\hat{a}_{\pm,vac}(t')$ in equations
\eqref{tmodeope} and \eqref{smodeope}, where $\eta_{t/s}$ is the
trigger/signal detector efficiency and $\hat{a}_{\pm,vac}$ are field
operators acting on vacuum.

With the single mode operators established we proceed as in the
second part of the last section, but since three modes are now
included, the covariance matrix $V$ is $6\times6$,
$y=(x_1,p_1,x_2,p_2,x_3,p_3)^T$, and $\pi^2$ is replaced by $\pi^3$
in Eq.\ \eqref{wvy}. To calculate the covariance matrix elements in
terms of $f_3(t)$, $\eta_t$, $\eta_s$, and OPO parameters we need
the two-time correlation functions for the nondegenerate OPO output.
These are \cite{drummond}
\begin{eqnarray}
\langle\hat{a}_\pm(t)\hat{a}_\mp(t')\rangle&=&
\frac{\lambda^2-\mu^2}{4}\left(\frac{\mathrm{e}^{-\mu|t-t'|}}{2\mu}+
\frac{\mathrm{e}^{-\lambda|t-t'|}}{2\lambda}\right),\nonumber\\
\langle\hat{a}^\dag_\pm(t)\hat{a}_\pm(t')\rangle&=&
\frac{\lambda^2-\mu^2}{4}\left(\frac{\mathrm{e}^{-\mu|t-t'|}}{2\mu}-
\frac{\mathrm{e}^{-\lambda|t-t'|}}{2\lambda}\right),\nonumber\\
\langle\hat{a}_\pm(t)\hat{a}_\pm(t')\rangle&=&
\langle\hat{a}^\dag_\pm(t)\hat{a}_\mp(t')\rangle=0,\label{correlation}
\end{eqnarray}
where $\lambda=\frac{\gamma}{2}+\epsilon$,
$\mu=\frac{\gamma}{2}-\epsilon$, $\epsilon$ is the nonlinear gain
coefficient of the OPO, and $\gamma$ is the OPO output mirror
leakage rate introduced above. Note that the dimension less twin
beam intensity
$\langle\hat{a}^\dag_\pm(t)\hat{a}_\pm(t)\rangle/\gamma$ is an
increasing function of $\epsilon/\gamma$.

In analogy to Eq.\ \eqref{Wclicks} the single-mode Wigner function
for the state conditioned on two trigger detector clicks is
\begin{multline}\label{Wclick}
W_{click}(x_3,p_3)=N_{click}\int\prod_{i=1}^2\bigg[dx_i
dp_i\frac{1}{2}\bigg(1+x_i^2+p_i^2\\
+x_i\frac{\partial}{\partial x_i}+p_i\frac{\partial}{\partial p_i}
+\frac{1}{4}\frac{\partial^2}{\partial x_i^2}
+\frac{1}{4}\frac{\partial^2}{\partial p_i^2}\bigg)\bigg]W_V(y)\\
=\frac{1}{C_1} \big[C_2+C_3(x_3^2+p_3^2)
+C_4(x_3^2+p_3^2)^2\big]\mathrm{e}^{-C_5(x_3^2+p_3^2)},
\end{multline}
where $C_1=D_1V_{55}\pi$, $C_2=D_1-V_{55}D_2$,
$C_3=D_2-2V_{55}V_{15}^2V_{35}^2$, $C_4=V_{15}^2V_{35}^2$,
$C_5=(V_{55})^{-1}$, and
\begin{eqnarray*}
D_1&=&V_{55}^4\left[(V_{11}-1)(V_{33}-1)+V_{13}^2\right],\\
D_2&=&V_{55}\big\{2V_{15}V_{35}\left(V_{13}V_{55}-V_{15}V_{35}\right)\\
&&+V_{55}\left[V_{15}^2(V_{33}-1)+V_{35}^2(V_{11}-1)\right]\big\}.\\
\end{eqnarray*}
$W_{click}$ is independent of $\eta_t$.

If the trigger beam is divided into $m$ beams, the field operator
representing the field in the $j$th beam may be written as
$c_{j0}\hat{a}_+(t)+\sum_{i=1}^{m-1}c_{ji}\hat{a}_{i,vac}(t)$, where
$c_{ji}$ are coefficients determined from the precise arrangement of
beam splitters, and $\hat{a}_{i,vac}(t)$ are field operators
representing vacuum states. If a click is observed in the $j$th and
the $k$th trigger beam in the temporal modes $f_1(t)$ and $f_2(t)$,
respectively, and we trace over all modes except the output mode
(denoted by $\mathrm{Tr}'$), the density operator $\rho_{tot}$ is
transformed into
\begin{multline}\label{rho}
\rho_{tot}\rightarrow \mathrm{Tr}' \Bigg[\int
f_1(t)\left(c_{j0}\hat{a}_+(t)+\sum_{i=1}^{m-1}c_{ji}\hat{a}_{i,vac}(t)\right)
dt\\
\int
f_2(t')\left(c_{k0}\hat{a}_+(t')+\sum_{i=1}^{m-1}c_{ki}
\hat{a}_{i,vac}(t')\right)dt'\\
\rho_{tot} \int f_2(t'')\left(c^*_{k0}\hat{a}^\dag_+(t'')
+\sum_{i=1}^{m-1}c^*_{ki} \hat{a}^\dag_{i,vac}(t'')\right)dt''\\
\int f_1(t''')\left(c^*_{j0}\hat{a}^\dag_+(t''')+
\sum_{i=1}^{m-1}c^*_{ji}\hat{a}^\dag_{i,vac}(t''')\right)dt'''\Bigg],
\end{multline}
where a normalization factor is omitted. The density operator
$\rho_{tot}$ is the direct product of the density operator for the
OPO output and the density operators for the vacuum states coupled
into the system via the beam splitters. Since the annihilation
operator acting on a vacuum state is zero, \eqref{rho} simplifies to
\begin{multline}
\rho_{tot}\rightarrow \mathrm{Tr}_{12} \Bigg[\int
f_1(t)c_{j0}\hat{a}_+(t)dt
\int f_2(t')c_{k0}\hat{a}_+(t')dt'\\
\rho_{123} \int f_2(t'')c^*_{k0}\hat{a}^\dag_+(t'')dt'' \int
f_1(t''')c^*_{j0}\hat{a}^\dag_+(t''')dt'''\Bigg],
\end{multline}
where the trace is now over the two trigger modes, and $\rho_{123}$
is the density operator for the two trigger modes and the output
mode. The factor $c_{j0}c_{k0}c^*_{k0}c^*_{j0}$ is irrelevant
because the transformed density operator has to be normalized. The
conditional output state is thus independent of the trigger detector
configuration, and it is justified to use the simple setup in figure
\ref{setup} in a theoretical treatment. One can verify that the
Wigner function of the conditional state continuously approaches the
outcome of a two-photon detection in a single trigger mode, when the
click separation approaches zero. The arguments are immediately
generalized to the case of $n$ conditioning clicks.

Finally, the two-photon state fidelity of the produced state is
obtained from the conditional Wigner function $W_{click}$ as in Eq.\
\eqref{F2}
\begin{multline}\label{tofid}
F_2(f_3(t))=\frac{2}{D_1(1+V_{55})^5}\big\{D_1(1-V_{55})^2(1+V_{55})^2\\
-D_2V_{55}^2(1-V_{55})(1+V_{55})(5-V_{55})\\
+2V_{55}^3(V_{15}V_{35})^2 [4V_{55}-5(1-V_{55})^2]\big\}.
\end{multline}
The fidelity depends on the choice of signal mode function $f_3(t)$.
In the next section we first determine the optimal mode function
$f_{op}(t)$, which leads to the largest fidelity, and then we
present explicit results for the predictions of equation
\eqref{tofid}.

\section{Optimal signal mode function and two-photon state fidelity}

In the low intensity limit \eqref{tofid} reduces to
\begin{equation}
\lim_{\epsilon\rightarrow0}F_2(f_3(t))=\frac{V_{15}^2V_{35}^2}
{2\left[(V_{11}-1)(V_{33}-1)+V_{13}^2\right]},
\end{equation}
and for small $\epsilon/\gamma$
\begin{eqnarray}
V_{15}&=&2\epsilon\sqrt{\Delta t_c\eta_t\eta_s}\int
f_3(t')\mathrm{e}^{-\frac{\gamma}{2}|t'-t_{c1}|}dt',\\
V_{35}&=&2\epsilon\sqrt{\Delta t_c\eta_t\eta_s}\int
f_3(t')\mathrm{e}^{-\frac{\gamma}{2}|t'-t_{c2}|}dt',
\end{eqnarray}
while $V_{11}$, $V_{33}$, and $V_{13}$ are always independent of
$f_3(t)$. The optimal signal mode function at very low intensity is
thus easily obtained by optimization of $V_{15}V_{35}$ under the
constraint $\int f_{op}(t)^2dt=1$. This leads to
\begin{multline}\label{tooutmodefunc}
\lim_{\epsilon\rightarrow0}f_{op}(t)=
N\bigg[\exp\left(-\frac{\gamma}{2}|t-t_{c1}|\right)\\
+\exp\left(-\frac{\gamma}{2}|t-t_{c2}|\right)\bigg],
\end{multline}
where $N$ is a normalization constant.

\begin{figure}
\begin{center}
\includegraphics*[viewport=8 5 387 297,width=0.95\columnwidth]{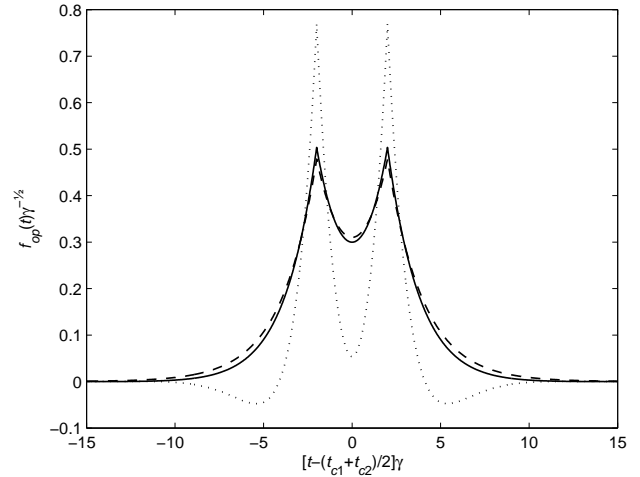}
\end{center}
\caption{Optimized mode functions for $|t_{c2}-t_{c1}|\gamma=4$ and
$\eta_s=1$. The three curves correspond to $\epsilon/\gamma=0$
(dashed line), $\epsilon/\gamma=0.08$ (solid line), and
$\epsilon/\gamma=0.2$ (dotted line).} \label{mode}
\end{figure}

For larger intensities the fidelity can be optimized numerically by
varying the shape of the mode function until no further increase in
fidelity is obtained. Optimized mode functions for
$|t_{c2}-t_{c1}|\gamma=4$ and different intensities are shown in
figure \ref{mode}. The peaks of the mode functions become sharper
with increasing intensity, and a dip appears on each side of the
peaks. This behavior is qualitatively the same as what was found for
the single peaked mode function of single-photon state generation in
\cite{nm}.

\begin{figure}
\begin{center}
\includegraphics*[viewport=19 9 387 295,width=0.95\columnwidth]{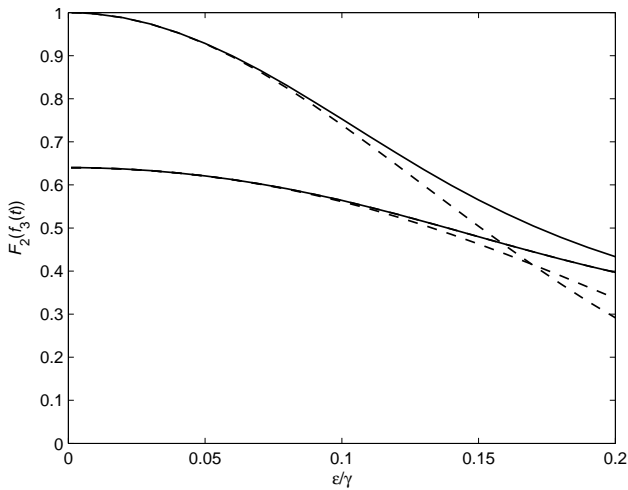}
\end{center}
\caption{Fidelity for $t_{c1}=t_{c2}$ calculated using the optimized
signal mode function (solid lines) and the optimal mode function at
zero intensity \eqref{tooutmodefunc} (dashed lines). Perfect signal
detection $\eta_s=1$ is assumed for the curves approaching 1 to the
left, while $\eta_s=0.8$ for the curves approaching 0.64.}
\label{toepsfid}
\end{figure}

Figure \ref{toepsfid} shows the optimized fidelity (solid lines) as
a function of $\epsilon/\gamma$ for $t_{c1}=t_{c2}$ and the fidelity
calculated from the optimal mode function at zero intensity (dashed
curves). The solid and dashed curves are almost identical for small
$\epsilon/\gamma$, so the optimal mode function at zero intensity is
close to optimal in the region where the fidelity is high and hence
provides an analytical approximation to the optimal choice of signal
state mode function. The figure shows that the fidelity decreases
when the intensity increases. This is as expected because a larger
mean photon flux results in larger contributions from higher photon
number states to the output state.

\begin{figure}
\begin{center}
\includegraphics*[viewport=15 5 387 295,width=0.95\columnwidth]{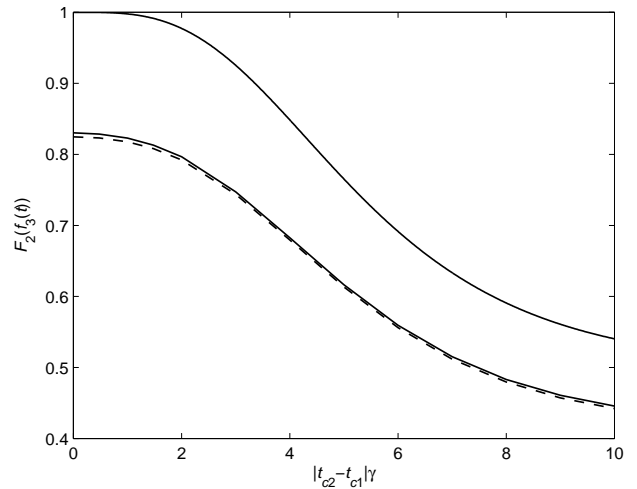}
\end{center}
\caption{Fidelity as a function of temporal distance between the
trigger detector clicks for $\epsilon/\gamma=0$ (upper solid line)
and $\epsilon/\gamma=0.08$ (lower solid line) calculated using the
optimized signal mode function. The dashed line shows the fidelity
for $\epsilon/\gamma=0.08$ obtained from the mode function
\eqref{tooutmodefunc}. Perfect signal detection $\eta_s=1$ is
assumed.} \label{antfid}
\end{figure}

The fidelity decreases when the temporal distance between the
trigger detector clicks increases from zero as is apparent from
figure \ref{antfid}, which shows the fidelity as a function of
$|t_{c2}-t_{c1}|\gamma$. When the click distance is different from
zero, the two photons no longer belong to precisely the same mode.
We return to this point in section V, where we also obtain an
analytical expression for the two-photon state fidelity as a
function of click distance in the small intensity limit. The
fidelity calculated from the optimal mode function at zero intensity
\eqref{tooutmodefunc} is also shown in figure \ref{antfid} (the
dashed curve), and it is seen that \eqref{tooutmodefunc} is a good
approximation to the optimal mode function even if the click
distance is large (provided $\epsilon/\gamma$ is not too large).

We note that the generation method described here favors close click
events since
\begin{multline}
\frac{\langle\hat{a}_+^\dag(t_{c1})\hat{a}_+^\dag(t_{c2})\hat{a}_+(t_{c2})
\hat{a}_+(t_{c1})\rangle}
{\langle\hat{a}_+^\dag(t_{c1})\hat{a}_+(t_{c1})\rangle
\langle\hat{a}_+^\dag(t_{c2})\hat{a}_+(t_{c2})\rangle}=\\
1+\left(\frac{\mathrm{e}^{-\mu|t_{c2}-t_{c1}|}}{2\mu}-
\frac{\mathrm{e}^{-\lambda|t_{c2}-t_{c1}|}}{2\lambda}\right)^2\bigg/
\left(\frac{1}{2\mu}-\frac{1}{2\lambda}\right)^2
\end{multline}
increases from one to two when $|t_{c2}-t_{c1}|\gamma$ decreases
from infinity to zero: the trigger events are bunched in time.

\section{Mode occupation description of the conditional state in the
low intensity limit}

In the previous sections we characterized the output state by the
Wigner function $W_{click}$ from which we were able to calculate the
fidelity for an arbitrary state in an arbitrary mode. However, a
deeper understanding of the nature of the conditionally produced
signal beam state can be obtained by considering the state as built
up of photons occupying specific temporal modes. A detailed mode
description of multi-photon states and manipulations of such states
is given in Ref.\ \cite{rohde}. In the present section we use this
approach to investigate the state generated when conditioning on $n$
trigger detector click events. It is assumed throughout that
$\epsilon/\gamma\ll1$.

If the trigger detector clicks are far apart, we know from Ref.\
\cite{nm} that the fidelity for a single photon in each of the $n$
modes
\begin{equation}\label{g}
g_i(t)\equiv\sqrt{\frac{\gamma}{2}}\mathrm{e}^{-\frac{\gamma}{2}|t-t_{ci}|},
\qquad i=1,2,\ldots,n
\end{equation}
is unity. On the other hand, for $n=2$ and $t_{c1}=t_{c2}$ we found
in the last section that the fidelity is unity for two photons in
the mode $g_1(t)$. In both limits the state generated in the signal
beam conditioned on two clicks is thus on the form
$|\psi_2\rangle=N_{\psi_2}\iint dtdt' g_1(t)g_2(t')\hat{a}_-^\dag(t)
\hat{a}_-^\dag(t')|0\rangle$, where $N_{\psi_2}$ is a normalization
constant. One is thereby led to consider whether this result is also
valid for intermediate separation of the trigger detector clicks. In
the appendix we show that the state generated in the signal beam
when conditioning on $n$ trigger detector click events is
\begin{equation}\label{state}
|\psi_n\rangle=N_{\psi_n}\left[\prod_{i=1}^n\int dt_i
g_i(t_i)\hat{a}_-^\dag(t_i)\right]|0\rangle,
\end{equation}
where $N_{\psi_n}$ is a normalization constant.

\begin{figure}
\includegraphics*[viewport=9 5 400 297,width=0.95\columnwidth]{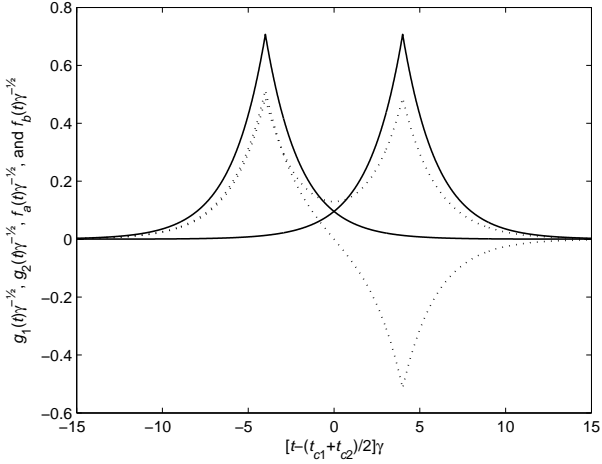}
\caption{Mode functions $g_1(t)$ and $g_2(t)$ (solid lines) and
$f_a(t)$ and $f_b(t)$ (dotted lines) for
$(t_{c2}-t_{c1})\gamma=8$.}\label{modefunc}
\end{figure}

\begin{figure}
\begin{center}
\includegraphics*[viewport=19 7 385 298,width=0.95\columnwidth]{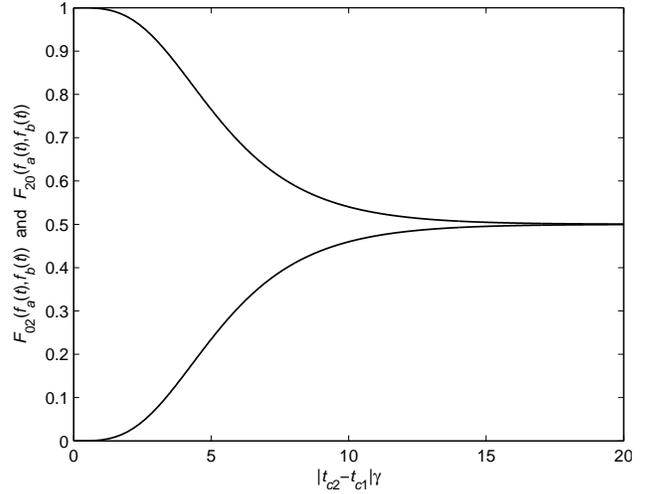}
\end{center}
\caption{The upper (lower) curve shows the probability to detect two
(zero) photons in the mode $f_a(t)$ and zero (two) photons in the
mode $f_b(t)$ as a function of temporal distance between trigger
detector clicks when $\epsilon/\gamma\rightarrow0$.}
\label{totfidab}
\end{figure}

To illustrate the meaning of \eqref{state} we consider the case
$n=2$ in some detail. Two orthogonal mode functions are constructed
from $g_1(t)$ and $g_2(t)$:
\begin{eqnarray}\label{fa}
f_a(t)&=&\frac{1}{\sqrt{2(1+I_{12})}}\left[g_1(t)+g_2(t)\right],\\
\label{fb}
f_b(t)&=&\frac{1}{\sqrt{2(1-I_{12})}}\left[g_1(t)-g_2(t)\right],
\end{eqnarray}
where
\begin{eqnarray}\label{overlap}
I_{ij}&\equiv&\int g_i(t)g_j(t)dt\nonumber\\
&=&\left(1+\frac{\gamma}{2}|t_{ci}-t_{cj}|\right)
\exp\left(-\frac{\gamma}{2}|t_{ci}-t_{cj}|\right).
\end{eqnarray}
The four mode functions are illustrated in figure \ref{modefunc} for
$\gamma(t_{c2}-t_{c1})=8$. Inserting \eqref{fa} and \eqref{fb} in
\eqref{state} we obtain
\begin{equation}\label{stateab}
|\psi_2\rangle=\frac{1+I_{12}}{\sqrt{2(1+I_{12}^2)}}|2,0\rangle_{ab}
-\frac{1-I_{12}}{\sqrt{2(1+I_{12}^2)}}|0,2\rangle_{ab},
\end{equation}
where $|x,y\rangle_{ab}=|x\rangle_a\otimes|y\rangle_b$ and
$|x\rangle_j$ means $x$ photons in the mode $f_j(t)$. Figure
\ref{totfidab} shows the norm square of the coefficients in
\eqref{stateab}. Since $f_a(t)$ is identical to the optimal mode
function for $\epsilon/\gamma\ll1$ \eqref{tooutmodefunc}, the upper
curve shows the maximal two-photon state fidelity as a function of
temporal distance between the conditioning clicks (i.e.\ the same
curve as in figure \ref{antfid}). For small $\gamma|t_{c2}-t_{c1}|$
we find from equations \eqref{overlap} and \eqref{stateab} that
\begin{equation}
F_2(f_a(t))\simeq1- \left(\frac{\gamma}{4}|t_{c2}-t_{c1}|\right)^4.
\end{equation}
Hence we still have a good two-photon state even if the trigger
detector clicks are not exactly simultaneous.

\begin{figure}
\begin{center}
\includegraphics*[viewport=10 6 370 298,width=0.95\columnwidth]{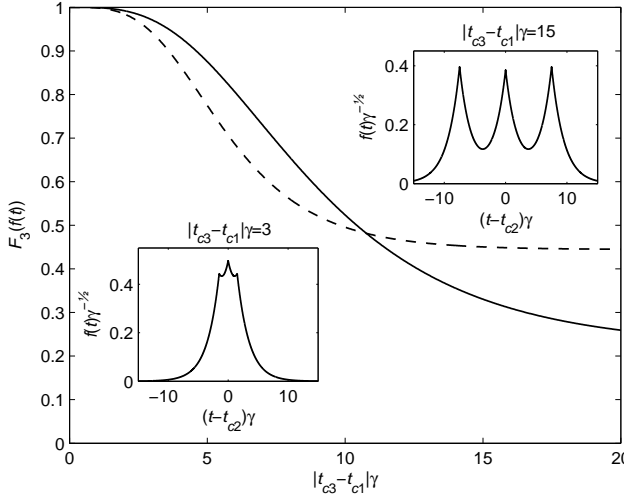}
\end{center}
\caption{Three-photon state fidelity as a function of temporal
distance between the first and the last trigger detector clicks for
$t_{c3}-t_{c2}=t_{c2}-t_{c1}$ (solid line) and $t_{c1}=t_{c2}$
(dashed line). The insets show the optimal mode function $f(t)$ for
the former case for $|t_{c3}-t_{c1}|\gamma=3$ (left) and
$|t_{c3}-t_{c1}|\gamma=15$ (right). $\epsilon/\gamma\ll1$ is
assumed.} \label{tF3}
\end{figure}

The optimal mode function for a general $n$ is obtained by
maximizing the fidelity between \eqref{state} and an $n$-photon
state
\begin{equation}
|n\rangle_f=\frac{1}{\sqrt{n!}} \left[\int
dtf(t)\hat{a}_-^\dag(t)\right]^n|0\rangle,
\end{equation}
where $f(t)$ is the mode function to be optimized. The $n$-photon
state fidelity is
\begin{equation}
F_n(f(t))=|\langle\psi_n|n\rangle_f|^2=n!|N_{\psi_n}|^2\prod_{i=1}^n\left|\int
f(t_i)g_i(t_i)dt_i\right|^2.
\end{equation}
It is apparent that the phase of $f(t)$ should be time independent
to maximize $F_n(f(t))$, and hence we choose $f(t)$ real.
Variational optimization of $\prod_{i=1}^n\int f(t_i)g_i(t_i)dt_i$
leads to
\begin{equation}
\xi f(t)=\sum_{i=1}^n g_i(t)\prod_{\substack{j=1 \\ j\neq i}}^n\int
f(t_j)g_j(t_j)dt_j,
\end{equation}
and thus
\begin{equation}
f(t)=\sum_{i=1}^n c_ig_i(t),
\end{equation}
where the constants $c_i$ and $\xi$ are determined from the highly
nonlinear set of equations
\begin{equation}\label{cs}
\xi c_i=\prod_{\substack{j=1 \\ j\neq i}}^n\sum_{k=1}^nc_kI_{kj}
\end{equation}
and
\begin{equation}\label{norm}
1=\sum_{i=1}^n\sum_{j=1}^nc_ic_jI_{ij}.
\end{equation}
For $n=2$ equations \eqref{cs} and \eqref{norm} are $\xi
c_1=c_1I_{12}+c_2$, $\xi c_2=c_1+c_2I_{12}$, and
$1=c_1^2+c_2^2+2c_1c_2I_{12}$, and hence
$c_1=c_2=1/\sqrt{2(1+I_{12})}$ and $\xi=1+I_{12}$ in agreement with
Eq.\ \eqref{tooutmodefunc}. For $n=3$ and
$t_{c2}-t_{c1}=t_{c3}-t_{c2}$ we obtain
\begin{eqnarray}
c_1&=&c_3=\sqrt{\frac{I_{12}^2-2(1+I_{13})+I_{12}\sqrt{I_{12}^2+4(1+I_{13})}}
{6[2I_{12}^2-(1+I_{13})](1+I_{13})}},\nonumber\\
c_2&=&-2c_1I_{12}+\sqrt{1+2c_1^2[2I_{12}^2-(1+I_{13})]},
\end{eqnarray}
and for $t_{c2}=t_{c1}$
\begin{eqnarray}
c_1&=&c_2=\sqrt{\frac{(4-I_{13}^2)-\sqrt{(4-I_{13}^2)^2-16(1-I_{13}^2)}}
{24(1-I_{13}^2)}},\nonumber\\
c_3&=&\frac{1-6c_1^2}{3c_1I_{13}}.
\end{eqnarray}
The three-photon state fidelity for these special cases is shown as
a function of click distance in figure \ref{tF3}. In the former case
the fidelity decreases from unity to $3!/3^3$, when the click
distance increases from zero to infinity, while in the latter case
it decreases from unity to $3\cdot2^2/3^3$, but in both cases a
broad region with a fidelity close to unity exists.

\section{Conclusion}
In conclusion we presented a theoretical description of conditional
higher photon number state generation from a continuous-wave light
source. We calculated the Wigner function for the output state
conditioned on two trigger detector clicks and determined its
overlap with a two-photon state. The output state mode function that
gives rise to the largest overlap was found. In the low intensity
limit, we showed that the state generated in the signal beam when
conditioning on $n$ trigger detector click events is
$|\psi_n\rangle=N_{\psi_n}\left[\prod_{i=1}^n\int dt_i
g_i(t_i)\hat{a}_-^\dag(t_i)\right]|0\rangle$, where $g_i(t)$ is a
function localized around the time for the $i$th click. From this we
obtain the optimal mode function and fidelity for $n$-photon state
generation at low intensity. For small temporal distance between the
trigger detector clicks and $n=2$ the optimized fidelity is unity
minus a small correction of fourth order in the temporal click
distance.

In the present treatment we averaged over all possible trigger
detector outcomes outside the small time windows specified by the
trigger mode functions, but for finite intensities larger fidelities
are obtained if we condition on dark intervals between the $n$
clicks (see \cite{nm}).

\appendix

\section{Conditional signal beam state at low intensity}
In this appendix we prove that the state generated in the signal
beam when conditioning on $n$ trigger detector clicks at times
$t_{c1},\ldots,t_{cn}$ is given by \eqref{state} in the limit
$\epsilon/\gamma\rightarrow0$. Since the normally ordered moments
determine the phase space P function \cite{lee}, it is sufficient to
prove that the expectation value obtained from \eqref{state} of all
normally ordered products of signal beam field operators is equal to
the expectation value obtained from the two-time correlation
functions \eqref{correlation}, when conditioning on the trigger
detector clicks, i.e.
\begin{multline}\label{exp}
\lim_{\epsilon\rightarrow0}\bigg\{\Big\langle
\Big[\prod_{i=1}^n\hat{a}_+^\dag(t_{ci})\Big]
\Big[\prod_{j=1}^m\hat{a}_-^\dag(t'_j)\Big]
\Big[\prod_{k=1}^p\hat{a}_-(t''_k)\Big]\\
\Big[\prod_{q=1}^n\hat{a}_+(t_{cq})\Big]\Big\rangle\Big/
\Big\langle\Big[\prod_{i=1}^n\hat{a}_+^\dag(t_{ci})\Big]
\Big[\prod_{q=1}^n\hat{a}_+(t_{cq})\Big]\Big\rangle\bigg\}=\\
\int\Big[\prod_{i=1}^ng_i(t_i)\Big]\Big[\prod_{q=1}^ng_q(t_{n+q})\Big]
\langle0|\Big[\prod_{i=1}^n\hat{a}_-(t_i)\Big]\\
\Big[\prod_{j=1}^m\hat{a}_-^\dag(t'_j)\Big]
\Big[\prod_{k=1}^p\hat{a}_-(t''_k)\Big]
\Big[\prod_{q=1}^n\hat{a}^\dag_-(t_{n+q})\Big]|0\rangle\\
\Big(\prod_{i=1}^{2n}dt_i\Big)
\Big/\int\Big[\prod_{i=1}^ng_i(t_i)\Big]
\Big[\prod_{q=1}^ng_q(t_{n+q})\Big]\\
\langle0|\Big[\prod_{i=1}^n\hat{a}_-(t_i)\Big]
\Big[\prod_{q=1}^n\hat{a}^\dag_-(t_{n+q})\Big]|0\rangle
\Big(\prod_{i=1}^{2n}dt_i\Big)
\end{multline}
The left hand side is evaluated using Wick's theorem for Gaussian
states with zero means \cite{louisell}, which states that
\begin{multline}\label{Wick}
\langle\hat{A}_1\hat{A}_2\ldots\hat{A}_{2k}\rangle=
\langle\hat{A}_1\hat{A}_2\rangle\langle\hat{A}_3\hat{A}_4
\hat{A}_5\ldots\hat{A}_{2k}\rangle\\+
\langle\hat{A}_1\hat{A}_3\rangle\langle\hat{A}_2\hat{A}_4
\hat{A}_5\ldots\hat{A}_{2k}\rangle +\ldots\\+
\langle\hat{A}_1\hat{A}_{2k}\rangle\langle\hat{A}_2\hat{A}_3
\hat{A}_4\ldots\hat{A}_{2k-1}\rangle,
\end{multline}
while $\langle\hat{A}_1\hat{A}_2\ldots\hat{A}_{2k+1}\rangle=0$,
where $\hat{A}_i$ is either an annihilation or a creation operator
and $k$ is a positive integer. The left hand side may thus be
expressed in terms of
$\langle\hat{a}^\dag_\pm(t)\hat{a}_\pm(t')\rangle$ and
$\langle\hat{a}_\pm(t)\hat{a}_\mp(t')\rangle$. For small
$\epsilon/\gamma$
\begin{eqnarray}\label{ada}
\langle\hat{a}^\dag_\pm(t)\hat{a}_\pm(t')\rangle&=&
\frac{2\epsilon^2}{\gamma}
\left(1+\frac{\gamma}{2}|t-t'|\right)\mathrm{e}^{-\frac{\gamma}{2}|t-t'|},\\
\label{aa} \langle\hat{a}_\pm(t)\hat{a}_\mp(t')\rangle&=&
\sqrt{\frac{2\epsilon^2}{\gamma}}
\sqrt{\frac{\gamma}{2}}\mathrm{e}^{-\frac{\gamma}{2}|t-t'|}.
\end{eqnarray}
It follows that the left hand side of \eqref{exp} is zero if $m\neq
p$, since in that case we have to combine operators, where the
expectation value of the product of the operators is zero. It also
follows that the denominator on the left hand side of \eqref{exp} is
proportional to $\epsilon^{2n}$. In the numerator the lowest order
terms in $\epsilon$ are obtained by combining as many
$\hat{a}^\dag_-$ operators as possible with $\hat{a}^\dag_+$
operators and as many $\hat{a}_-$ operators as possible with
$\hat{a}_+$ operators. These terms are proportional to
$\epsilon^{\max(2n,2m)}$ for $m=p$. To obtain a nonzero left hand
side in the limit $\epsilon/\gamma\rightarrow0$, it is thus
necessary to require that $m=p\leq n$. It is immediately apparent
that the right hand side of \eqref{exp} is zero if $m=p\leq n$ is
not satisfied.

We now consider the case $m=p\leq n$ and by combining the operators
as described above and using \eqref{aa} and \eqref{ada} and the
definitions \eqref{g} and \eqref{overlap} we obtain for the lowest
order terms of the left hand side (LHS) of \eqref{exp}
\begin{multline}\label{LHS}
\mathrm{LHS}=\frac{1}{(n-m)!}\sum_{P_i}\sum_{P_j}g_{i_1}(t'_1)\ldots
g_{i_m}(t'_m)g_{j_1}(t''_1)\ldots\\
g_{j_m}(t''_m)I_{i_{m+1},j_{m+1}}\ldots I_{i_n,j_n}\Big/\sum_{P_j}
I_{1,j_1}\ldots I_{n,j_n},
\end{multline}
where the summations are over all permutations of the $n$
$i$-indices and all permutations of the $n$ $j$-indices,
respectively. The right hand side of \eqref{exp} is evaluated using
the relation
\begin{multline}
\hat{a}_-(t''_k)\left[\prod_{i=1}^n\hat{a}_-^\dag(t_{i})\right]|0\rangle=\\
\sum_{j=1}^n\delta(t''_k-t_{j}) \left[\prod_{\substack{i=1
\\ i\neq j}}^n\hat{a}_-^\dag(t_{i})\right]|0\rangle
\end{multline}
repeatedly, which leads to
\begin{multline}\label{delta}
\langle0|\prod_{i=1}^n\hat{a}_-(t_i)
\prod_{j=1}^m\hat{a}_-^\dag(t'_j)\prod_{k=1}^m\hat{a}_-(t''_k)
\prod_{q=1}^n\hat{a}^\dag_-(t_{n+q})|0\rangle=\\
\frac{1}{(n-m)!}\sum_{P_i}\sum_{P_j}\delta(t'_1-t_{i_1})\ldots
\delta(t'_m-t_{i_m})\\
\delta(t''_1-t_{n+j_1})\ldots\delta(t''_m-t_{n+j_m})
\delta(t_{i_{m+1}}-t_{n+j_{m+1}})\\
\ldots\delta(t_{i_n}-t_{n+j_n}).
\end{multline}
Inserting \eqref{delta} in \eqref{exp} we immediately obtain the
result \eqref{LHS} for the right hand side of \eqref{exp}.

\end{document}